\documentclass[prl
,twocolumn%
,showpacs%
,aps]{revtex4}
\usepackage{natbib}
\usepackage{graphicx}
\begin{document}
\title{A Deterministic Single-Photon Source for Distributed Quantum Networking}
\author{Axel Kuhn, Markus Hennrich, and Gerhard Rempe}
\affiliation{Max-Planck-Institut f\"{u}r Quantenoptik, Hans-Kopfermann-Str.1, 85748 Garching, Germany}
\date{\today}
\begin{abstract}
A sequence of single photons is emitted on demand from a single three-level atom strongly 
coupled to a high-finesse optical cavity. The photons are generated by
an adiabatically driven stimulated Raman transition between two atomic ground states, 
with the vacuum field of the cavity stimulating one branch of the transition, and
laser pulses deterministically driving the other branch. This process is 
unitary and therefore intrinsically reversible, which is essential for  
quantum communication and networking, and the 
photons should be appropriate
for all-optical quantum information processing.
\end{abstract}
\maketitle

A future quantum network connecting remote quantum processors 
and memories has several advantages in processing quantum information 
as compared to a local quantum computer, 
since it combines 
scalability with modularity. Different kinds 
of networks have been proposed 
\cite{Monroe02}: 
one is an all-optical network 
\cite{Knill01}, 
where the nodes are linear optical components, with quantum information 
encoded in the number of photons flying from node to node.
The nodes perform gate operations based on quantum interference effects between 
indistinguishable photons. In another, more general, network 
the nodes also serve as quantum memories storing information, e.g., in 
long-lived states of atoms located in an optical cavity 
\cite{Cirac97}. 
The key requirement for
such a network is its ability to interconvert stationary and flying qubits
and to transmit flying qubits between specified locations
\cite{DiVincenzo00}. 
The atom-cavity system, in particular, must be able to transfer quantum information between 
atoms and photons in a coherent manner 
\cite{Maitre97,Brattke01}. 
It must also act as an emitter and a receiver
of single-photon states. These states must therefore be generated 
by a reversible process. However, all 
deterministic single-photon emitters demonstrated so far 
\cite{Kim99,Lounis00,Kurtsiefer00,Brouri00,Michler00,Santori01,Yuan02,Michler00:2,Moreau01}
do not meet this essential requirement. The reason is that the 
emission process, namely an electronic excitation of the system 
followed by spontaneous emission, cannot be described by a Hamiltonian 
evolution and, hence, is irreversible.

This letter describes the realization of an intrinsically reversible  single-photon source 
\cite{Cirac97,Enk97:1,Parkins93,Parkins95,Pellizari95}, 
which is based on a stimulated Raman process driving an adiabatic 
passage 
\cite{Vitanov01}
(STIRAP) between two ground states of a single atom strongly 
coupled to a single mode of a high-finesse optical cavity 
\cite{Law97,Kuhn99}. 
A laser beam illuminating the atom excites 
one branch of the Raman transition, while the cavity vacuum stimulates 
the emission of the photon on the other branch. STIRAP is slow 
compared to the photon lifetime in the cavity, so that the
field generated inside the cavity is instantaneously mapped to 
the outside world. Moreover, it employs a dark state, which has 
two important consequences: first, any electronic excitation 
is avoided, so that irreversible spontaneous processes 
do not occur. Second, the scheme allows one to continuously tune the frequency of the photon within 
a range that is only limited by the atom-cavity coupling strength. 
The tuning ability has recently been demonstrated with a beam 
of atoms passing through the cavity 
\cite{Hennrich00}. 
This experiment produced at most one photon per passing 
atom, but did not operate as a 
single-photon source, because its continuous driving scheme simply 
mapped the random (Poissonian) atom statistics to the photons.
The present experiment, however, uses a pulsed driving 
together with a pulsed recycling. This makes possible to produce 
on demand a stream of several single-photon pulses from one-and-the-same 
atom, triggered by the detection of a ``first'' photon emitted 
from the cavity. 

\begin{figure}
\includegraphics[width=8.5cm]{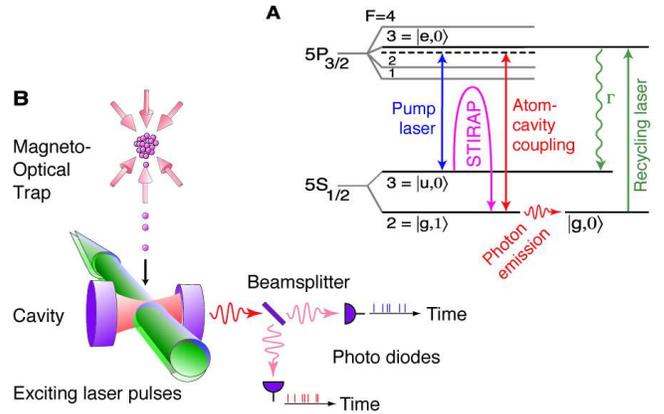}
\caption{\label{kfig1}Scheme of the experiment. \textbf{(A)} Relevant energy 
levels and transitions in $^{85}$Rb. The atomic states labeled 
$\left| u\right\rangle$, 
$\left| e\right\rangle $
 and 
$\left| g\right\rangle $
 are involved in the Raman process, and the states 
$\left| 0\right\rangle $
 and 
$\left| 1\right\rangle $
 denote the photon number in the cavity. \textbf{(B)} Setup: A cloud 
of atoms is released from a magneto-optical trap and falls through 
a cavity $20\,$cm below in about $8\,$ms with a velocity 
of $2\,$m/s. The interaction time of a single atom with the 
TEM$_{00}$ mode of the cavity (waist $w_{0}=35\,\mu$m) 
amounts to about $17.5\,\mu$s. The pump and recycling 
lasers are collinear and overlap with the cavity mode. Photons 
emitted from the cavity are detected by a pair of photodiodes 
with a quantum efficiency of 50\%.
}
\end{figure}

Figure \ref{kfig1}A shows the basic scheme of the photon-generation process. 
A single $^{85}$Rb atom is prepared in state 
$\left| u\right\rangle$,
which is the $F=3$  hyperfine state of the $5S_{1/2}$ electronic ground state. 
The atom is located in a high-finesse optical 
cavity, which is near resonant with the 
780\,nm transition between states 
$\left| g\right\rangle $
 and 
$\left| e\right\rangle$. 
Here, 
$\left| g\right\rangle $
 is the $F=2$ hyperfine state of the electronic ground state 
and 
$\left| e\right\rangle $
 is the electronically excited $5P_{3/2}(F=3)$ state. The state of the cavity 
is denoted by 
$\left| n\right\rangle$, 
where $n$ is the number of photons. When the atom is placed 
inside the cavity, the product states 
$\left| g,n\right\rangle $
 and 
$\left| e,n-1\right\rangle $
 are coupled by the electric dipole interaction, characterized 
by the Rabi frequency 
$\Omega _{n}^{} =2g\sqrt{n}$. 
Here, $g$ is the average atom-cavity coupling constant, which 
takes into account that neither the position of the atom in the 
cavity, nor the magnetic quantum number of the atom is well defined 
in the experiment. We assume $g$ to be constant while a pump-laser 
pulse with Rabi frequency 
$\Omega _{P} \left( t\right) $
 is applied. This laser is close to resonance with the 
$\left| u\right\rangle \leftrightarrow \left| e\right\rangle $
 transition, so that now the three product states 
$\left| u,n-1\right\rangle$, 
$\left| e,n-1\right\rangle $
 and 
$\left| g,n\right\rangle $
 of the atom-cavity system are coupled. For the one-photon manifold, 
$n=1$, 
and a Raman-resonant excitation, where the detunings of the 
pump laser, 
$\Delta _{P}$, and the cavity, 
$\Delta _{C}$, 
from the respective atomic transitions are equal, it is straightforward 
to find the three eigenstates of the coupled atom-cavity system, 
$\left| \phi _{1}^{\pm } \right\rangle $
 and 
$\left| \phi _{1}^{0} \right\rangle =\left[ 2g\left| u,0\right\rangle
-\Omega _{P} (t)\left| g,1\right\rangle \right] /\sqrt{4g^{2} +\Omega
_{P}^{2} (t)} $.
Note that state 
$\left| \phi _{1}^{0} \right\rangle $
 is dark, i.e. has no contribution of the excited state, 
$\left| e\right\rangle$, 
and is therefore not affected by spontaneous emission. 

The dark state 
$\left| \phi _{1}^{0} \right\rangle $
 is now used to generate a single photon inside the cavity. This 
is achieved by establishing a large atom-cavity coupling constant, $g$, 
before turning on the pump pulse. In this case, the system's 
initial state, 
$\left| u,0\right\rangle$, 
coincides with 
$\left| \phi _{1}^{0} \right\rangle$. 
Provided the pump pulse rises slowly, the system's state vector 
adiabatically follows any change of 
$\left| \phi _{1}^{0} \right\rangle$, 
and for a lossless cavity a smooth transition from 
$\left| u,0\right\rangle $
 to 
$\left| g,1\right\rangle $
 is realized as soon as 
$\Omega _{P} \gg 2g$. 
Hence, a single photon is generated in the relevant cavity 
mode.
This photon leaves the cavity through that mirror which is designed 
as an output coupler. The emission starts as soon as the decaying 
state, 
$\left| g,1\right\rangle$, 
contributes to 
$\left| \phi _{1}^{0} \right\rangle$, 
i.e. already with the rising edge of the pump pulse, because 
the contribution from 
$\left| g,1\right\rangle $
 is proportional to 
$\Omega _{P}^{2} (t)$. 
If the pump pulse rises slowly, the emission can therefore 
end even before 
$\Omega _{P} >2g$. 
The dynamics of the simultaneous excitation and emission processes 
determines the duration and, hence, the linewidth of the photon.
When the photon is emitted, the final state of the coupled 
system, 
$\left| g,0\right\rangle $, 
is reached. This state is not coupled to the one-photon manifold, 
and the atom cannot be reexcited. This limits the number of photons 
per pump pulse and atom to one.

To emit a sequence of photons from one-and-the-same atom, the 
system must be transferred back to 
$\left| u,0\right\rangle $
 once an emission has taken place. To do so, we apply recycling 
laser pulses that hit the atom between consecutive pump pulses. 
The recycling pulses are resonant with the 
$\left| g\right\rangle \leftrightarrow \left| e\right\rangle $
 transition and pump the atom to state 
$\left| e\right\rangle$. 
From there, it decays spontaneously to the initial state, 
$\left| u\right\rangle$. 
Note that state 
$\left| e\right\rangle $
 populated by the recycling laser couples to the cavity. However, 
spontaneous emission into the cavity is suppressed by deliberately 
choosing a large cavity detuning, 
$\Delta _{C}$. 
The pump laser is detuned by the same amount to assure Raman 
resonance. If an atom that resides in the cavity is now exposed 
to a sequence of laser pulses, which alternate between triggering 
single-photon emissions and re-establishing the initial condition 
by optical pumping, a sequence of single-photon pulses is produced.

\begin{figure}
\includegraphics[width=7.8cm]{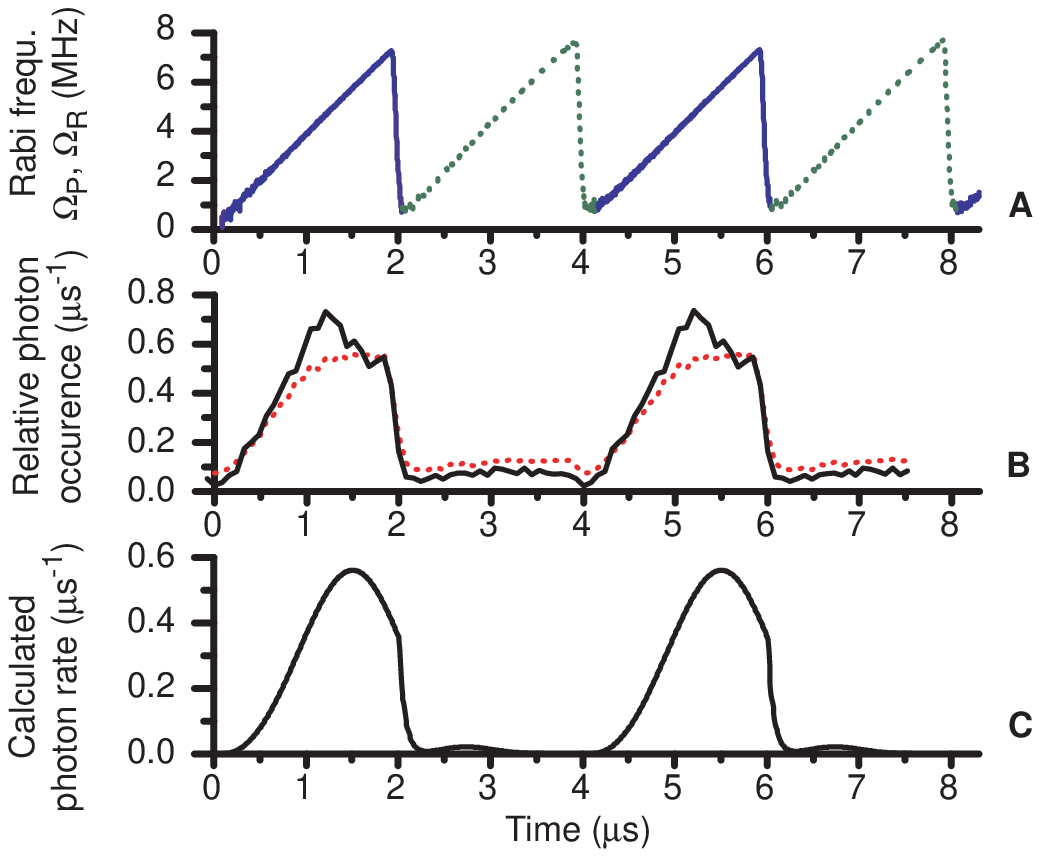}
\caption{\label{kfig2}Pulse shapes. \textbf{(A)} The atoms are periodically illuminated 
with $2\,\mu$s-long pulses from the pump (solid line) 
and the recycling laser (dotted line). \textbf{(B)} Measured arrival-time 
distribution of photons emitted from the cavity (dotted line). 
The solid line shows the arrival-time distribution of photons 
emitted from strongly coupled atoms (see text). \textbf{(C)} Simulation 
of the process with 
$(g,\;\Omega _{P,R}^{0} ,\;\Delta _{P,C} ,\;\Gamma ,\;\kappa )=2\pi \times
(2.5,\;8.0,\;-20.0,\;6.0,\;1.25)\,$MHz, 
where 
$\Omega _{P,R}^{0} $
 are the peak Rabi frequencies of the pump- and recycling pulses, 
and 
$\Gamma $
 and 
$\kappa $
 are the atom and cavity-field decay rates, respectively.}
\end{figure}

Figure \ref{kfig1}B shows the apparatus. Atoms are released from a magneto-optical 
trap and pass through the TEM$_{00}$ mode of the optical cavity, 
where they are exposed to the sequence of laser pulses. 
On average, 3.4 atoms/ms enter the cavity
\footnote{The flux of atoms is determined by a statistical analysis of the emitted light, with continuous pump- and recycling lasers exciting the falling atoms. As the cavity acts as detector,  non-interacting atoms are ignored, so that its spatial mode structure is taken into account.},
so that the probability of finding a single atom inside the cavity is 5.7\%, while the 
probability of having more than one atom is only 0.18\% which 
is negligible. The cavity is 1 mm long and has a finesse of $60\,000$. 
One mirror has a 25 times larger transmission coefficient than 
the other. Therefore, photons are preferentially emitted into 
one direction. These photons are counted by two avalanche photodiodes 
which are placed at the output ports of a beam splitter. For 
each experimental cycle, all photon arrival times are recorded 
with transient digitizers with a time resolution of $8\,$ns.

\begin{figure}
\includegraphics[width=7.5cm]{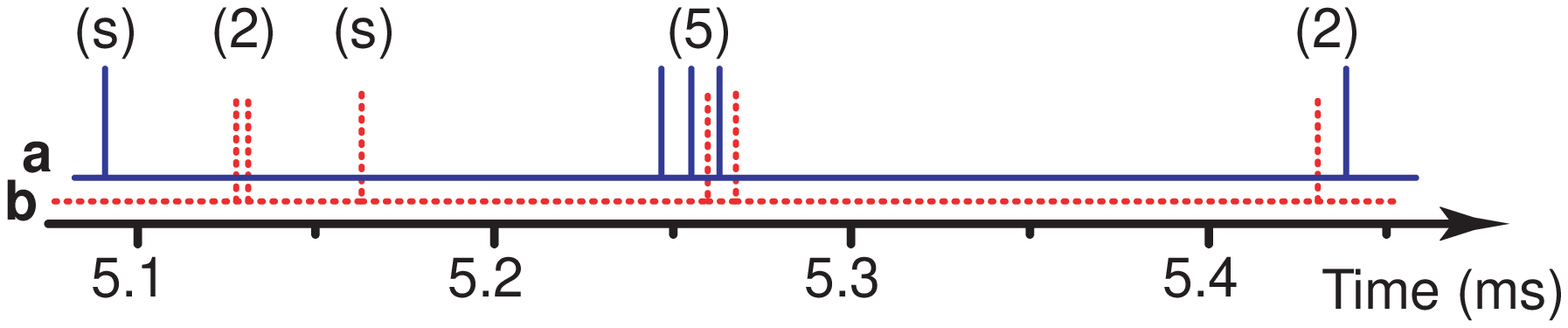}
\caption{\label{kfig3}Photon sequence: Clip of the photon streams 
arriving at the photodiodes D1 and D2 (traces \textbf{a} and \textbf{b}, 
respectively). Several sequences of two \textbf{(2)} and five \textbf{(5)} 
photon emissions are observed, with durations comparable to the 
atom-cavity interaction time. The solitary events \textbf{(s)} are 
either dark counts, or, more likely, photons coming from atoms 
that are only weakly coupled to the cavity.}
\end{figure}

In the experiment, the electric field amplitudes and, hence, 
the Rabi frequencies of the pump and recycling pulses have the 
shape of a saw-tooth and increase linearly, as displayed in Fig.\,\ref{kfig2}A. 
This leads to a constant rate of change of the dark state, 
$\left| \phi _{1}^{0} \right\rangle$, 
during the initial stage of the pump pulses, and therefore 
optimal adiabaticity with minimal losses to the other eigenstates. 
The linear slope of the recycling pulses suppresses higher Fourier 
components and therefore reduces photon emission into the detuned 
cavity. Note that the recycling process is finished before the 
end of the pulse is reached, so that the final sudden drop in 
Rabi frequency does not influence the atom. 

Also shown in Fig.\,\ref{kfig2} are two measured arrival-time distributions 
of the photons and a simulation of the photon emission rate for 
typical experimental parameters. The simulation is based on a 
numerical solution of the system's master equation 
\cite{Kuhn99}
which takes into account the decay of the relevant states. The 
simulation (Fig.\,\ref{kfig2}C) reveals that the pump-pulse duration of 
$2\,\mu$s is slightly too short, as the emitted photon 
pulse is not completely finished. This is also observed in the 
photon arrival-time distribution (Fig.\,\ref{kfig2}B). Here, the measured 
data agree well with the simulation if only photons from strongly coupled atoms 
are considered (solid line). For these, we assume that several photons are detected  within
he atom-cavity interaction time. 
If solitary photons, which we attribute to weakly coupled atoms, 
are included in the analysis, the arrival-time distribution 
is given by the dotted line. 
Note that the envelope of the photon pulses is well explained by the 
expected shape of the single-photon wavepackets, and therefore cannot be attributed 
to an uncertainty in emission time, which is not present for a unitary process. 
Assuming transform-limited Gaussian pulses, we infer a single-photon linewidth of
$\Delta\nu=340\,$kHz (FWHM) from the $1.3\,\mu$s photon-pulse duration (FWHM).
We emphasize that the pump-pulse 
duration was adjusted to maximize the number of photons per atom. 
Longer pump pulses would not truncate the photon pulses and, 
hence, would slightly increase the emission probability per pulse, 
but due to the limited atom-cavity interaction time, the total 
number of photons per atom would be reduced.

\begin{figure}
\includegraphics[width=7.5cm]{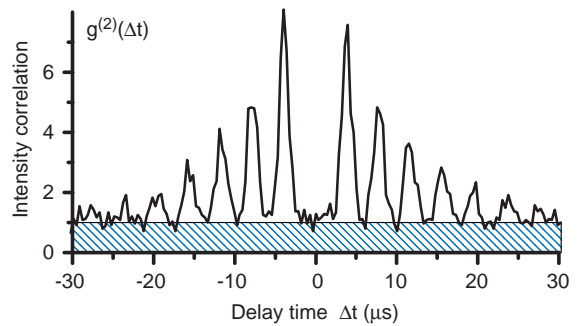}
\caption{\label{kfig4}Second-order intensity correlation 
of the emitted photon stream, averaged over 15000 experimental cycles
 (loading and releasing of the atom cloud) 
with a total number of 184868 photon counts. The hatched area represents correlations between 
photons and detector-noise counts.}
\end{figure}

Figure \ref{kfig3} displays an example of the photon stream recorded while 
single atoms fall through the cavity one after the other. Obviously, 
the photon sequence is different for each atom. In particular, 
not every pump pulse leads to a detected photon, since the efficiencies 
of photon generation and photon detection are limited. The second-order 
intensity correlation function of the emitted photon stream is 
shown in Fig.\,\ref{kfig4}. Displayed is the cross-correlation of the photon 
streams registered by the two photodiodes $D1$ and $D2$. It is defined 
as 
$g^{(2)} (\Delta t)=\left\langle P_{D1} (t)P_{D2} (t-\Delta
t)\right\rangle /\left( \left\langle P_{D1} (t)\right\rangle \left\langle
P_{D2} (t)\right\rangle \right) $, 
where $P_{D1}(t)$ and $P_{D2}(t)$ are the probabilities 
to detect a photon at time $t$ with photodiode $D1$ and $D2$, respectively. 
Note that all photon-arrival times are recorded to calculate 
the full correlation function, without the otherwise 
usual restriction of a simple start/stop measurement which would 
consider only neighboring events. Of course, 
$g^{(2)} $
 includes not only correlations between photons emitted from 
the cavity but also those involving detector-noise counts. This 
last contribution has been determined from an independent measurement 
of the detector-noise count rate. The result is indicated by 
the time-independent hatched area in Fig.\,\ref{kfig4}. Only the excess 
signal, 
$\tilde{g} ^{(2)} (\Delta t)=g^{(2)} (\Delta t)-g_{noise}^{(2)} $, 
reflects the true photon statistics of the light emitted from 
the atom-cavity system.

The correlation function, 
$\tilde{g} ^{(2)} (\Delta t)$, 
oscillates with the same periodicity as the sequence of pump 
pulses. This indicates that photons are only emitted during the 
pump pulses, and no emissions occur when recycling pulses are 
applied. The nearly Gaussian envelope of the comb-like function 
is obviously a consequence of the limited atom-cavity interaction 
time. The most remarkable feature in Fig.\,\ref{kfig4} is the missing correlation 
peak at 
$\Delta t=0$. 
In fact, photon antibunching together with 
$\tilde{g} ^{(2)} (0)\approx 0$
 is observed. This clearly demonstrates the nonclassical character 
of the emitted light, and proofs that (a) the number of emitted 
photons per pump pulse is limited to one, and (b) no further 
emission occurs before the atom is recycled to its initial state. 
Note that the relatively large noise contribution is no intrinsic 
limitation of our system but reflects only the low 
atomic flux through the cavity in the present experiment. 

We emphasize that the detection of a first photon signals the presence of an atom  
in the cavity, and fixes the atom number to one. The photons emitted from this atom
during subsequent pump pulses dominate the photon statistics and give
rise to antibunching. Such an antibunching would not be observed for faint laser pulses, 
since a random photon statistics applies to each pulse. 
The areas of the different peaks of the correlation function in Fig.\,\ref{kfig4}
reflect the probability for the emission of further photons from 
one-and-the-same atom. 
They are
determined from a lengthy but straightforward 
calculation, which relates the number of 
correlations per pulse with the total number of 
photons. Using the 
data displayed in Fig.\,\ref{kfig4}, the result for the conditional 
emission of another photon during the (next, 3$^{rd}$, 4$^{th}$, 5$^{th}$, 
6$^{th}$, 7$^{th}$) pump pulse is (8.8, 5.1, 2.8, 1.4, 0.8, 0.5)\,\%. 
Note that the probabilities for subsequent emissions decrease, 
since the photon emission probability, 
$P_{emit}$, 
depends on the location of the moving atom. It is highest for 
an atom in an antinode, and decreases if the atom moves away 
from this point. It is not possible to control the atoms location 
in the present experiment, but it is possible to calculate 
$P_{emit} (z)$
 from the experimental data. Here, 
$z$
 is the atom's vertical position relative to the cavity axis, 
and 
$P_{emit} (z)$
 is averaged over all possible atomic trajectories in the horizontal $xy$-plane. 
Assuming a Gaussian 
$z$-dependence, the deconvolution of 
$\tilde{g} ^{(2)} (\Delta t)$
gives 
$P_{emit} (z)=0.17\exp \left[ -\left( z/15.7\,\mu m \right) ^{2} \right] $.
For $z=0$, the average photon-emission probability of 17\% is smaller than the calculated value of 67\% 
for an atom in an antinode of the cavity. It follows that a system combining a cavity and a single atom at rest in a dipole trap \cite{Schlosser01,Kuhr01}, or a single ion at rest in a 
rf-trap \cite{Guthoerlein01,Mundt02}, 
should allow one to generate a continuous bit-stream 
of single photons with a large and time-independent efficiency
\cite{Law97,Kuhn99}. 
The photon repetition rate is limited by the atom-cavity coupling constant, $g$, which one
could push into the GHz regime by using smaller cavities of
wavelength-limited dimensions in, e.g., a photonic bandgap material.

In conclusion, we have shown that a coupled atom-cavity system 
is able to emit single photons on demand. Moreover, it is possible to generate a sequence of 
up to seven photons on demand from one-and-the-same atom in a 
time interval of about $30\,\mu$s. These photons are 
all  generated in a well-defined radiation mode. They should have 
the same frequency
and a Fourier-transform limited linewidth, 
limited from above by the decay rate of the cavity field
\cite{Hennrich00}. 
It follows 
that one can expect the photons to be indistinguishable and, 
therefore, ideal for all-optical quantum computation schemes 
\cite{Knill01}.
Moreover, the photon-generation process is unitary. This makes 
possible to produce arbitrarily shaped single-photon pulses by 
suitably tailoring the envelope of the pump pulse. For symmetric 
pulses, the emission process can be reversed. This should allow 
one to transfer the photon's quantum state to another atom located 
in another cavity. Such a state mapping between atoms and photons 
is the key to quantum teleportation of atoms between distant 
nodes in a quantum network of optical cavities 
\cite{Cirac97}.

\begin{acknowledgements}
This work was 
supported by the 
research program `Quantum Information Processing' of the Deutsche Forschungsgemeinschaft, 
and by the European Union through the IST (QUBITS) and IHP (QUEST) 
programs.
\end{acknowledgements}


\end{document}